# Photo-induced pair production and strong field QED on Gemini
## Initial experiment of a multi-part campaign

CH Keitel[1], A Di Piazza[1], GG Paulus[2], T Stöhlker[2], E Clark[3], S Mangles[4], Z Najmudin[4], K. Krushelnick[5], M Borghesi[6], B.Dromey[6], M. Geissler[6], D Riley[6], G Sarri[6] and M Zepf[2,6]

**Summary:**

The extreme laser intensities made possible by cutting edge laser systems such as Astra Gemini are opening up the possibility to reach, in the laboratory, conditions in which processes predicted by the non-linear Quantum Electrodynamics (QED) theory will be observable for the first time. Electron-positron pair production from a pure vacuum target is possibly the most iconic process that this theory predicts. However, despite its intrinsic fascination and central relevance in modern physics, this phenomenon is still to be experimentally observed. In addition to being a direct confirmation of a fundamental tenet of QED such an observation would represent an impressive demonstration of the equivalence of mass and energy. Beyond pair-production our campaign will allow the experimental investigation of currently unexplored extreme radiation regimes, like the quantum radiation dominated regime, where quantum and self-field effects dominate the electron dynamics.

The Astra Gemini laser provides the unique combination of two Petawatt class laser beams and therefore allows the interaction of secondary GeV electron and γ-ray beams with an ultra-intense laser focus with intensities in excess of $10^{21}$ Wcm$^{-2}$. Our calculations show that such conditions put detectable pair production from vacuum within reach. The complexity of the experiment – combining state-of-the-art laser accelerators with ultra-intense laser interactions – requires a staged approach in terms of its realisation. This proposal is for the first experiment in a series of 3 to achieve our high-profile experimental goal.

As well as being an essential stepping-stone to the first demonstration of non-linear QED effects like pair production in vacuum, each experiment in the series will have direct and tangible high-profile outcomes in terms of determining high brightness MeV - GeV electron and γ-ray fluxes, optimisation of the flux and stability of GeV electron beams, pair-production via electron-laser interaction (Trident and two-step Breit-Wheeler) and non-linear Compton scattering in the quantum radiation dominated regime.

**The scientific team:**

An international scientific team has been assembled to provide the mix of experimental and theoretical expertise required for a sustained and high profile campaign. The collaborating institutions are the Max-Planck-Institutes in Heidelberg and Munich, the Helmholtz Institute in Jena, Queen's University of Belfast (experimental coordination), Imperial College, Univ. of Michigan and TEI Crete. Together these groups have a formidable skill set spanning intense laser interactions and laser based electron accelerators, detector development and laser and QED theory – thus providing the full complement of highest level expertise required. Both the Helmholtz Institute and QUB have specifically funded programmes (through the Helmholtz foundation and an EPSRC Platform Grant specific to these goals) to pursue the new horizons of strong field QED.

**Background:**

Pair production from the vacuum is predicted to be possible either by the application of a (quasi-) static field of the order of the critical Schwinger limit of $E_{crit}\sim 10^{16}$ V/cm (so called 'vacuum breakdown') [1], or by the interaction of two counter-propagating photons with energies $\hbar\omega_1$ and $\hbar\omega_2$ such that $(\hbar\omega_1 \hbar\omega_2)^{1/2} > 2m_e c^2$, with $m_e c^2$ being the electron rest energy (γγ or Breit-Wheeler reaction [2,3]). That such processes of converting pure field energy to mass should exist is fundamental to our understanding of the interaction of fields, matter and vacuum. To date however, there has been no direct experimental verification of this fundamental process in vacuum. This is primarily due to both the extremely large value of the critical field $E_{crit}$ and the low collision cross sections of γ-rays, which demand extremely high photon fluxes.

Despite these challenging requirements our collaboration has identified what we believe to be a viable process to observe pair creation directly in a particle-free vacuum for the first time with lasers such as Astra Gemini. The concept relies on the collision of a γ-ray beam (produced by Bremsstrahlung from laser-accelerated electron) with an intense optical field (electric field amplitude E and angular frequency $\omega_{opt}$) provided by f/1 focusing of Gemini's second beam.

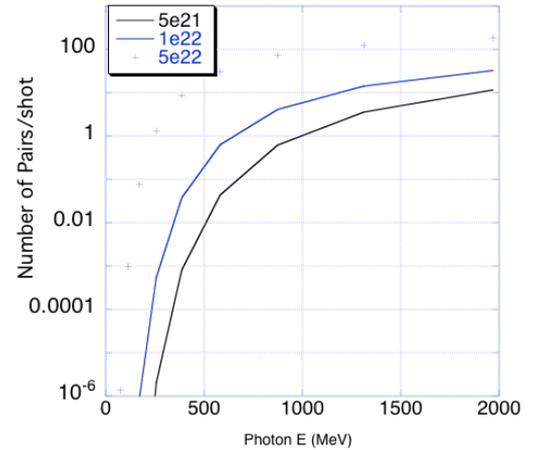

Fig. 1: Pair production in vacuum predicted per shot for a Bremsstrahlung γ-ray beam interacting with an intense laser beam for different laser intensities and laser pulse duration of 40 fs.

Under these circumstances pair production can be achieved via the Breit-Wheeler reaction if a single energetic γ-ray with energy $\hbar\omega_\gamma$ interacts with N optical photons (energy $\hbar\omega_{OPT}$) with a collision energy above the threshold: $(\hbar\omega_\gamma N\hbar\omega_{OPT})^{1/2} > 2m_e c^2$. Similar to strong field ionisation, this process can be achieved in two different regimes – multi-photon and tunnelling (or quasistatic). These regimes are identified by distinct scalings that depend on the laser strength parameter $a_0=|e|E/m_e\omega_{opt}c$ of the laser field.

If $a_0<1$ pair production occurs in the so-called perturbative or multi-photon regime and pair-production scales with the N-th power of the laser intensity. On the other hand, if $a_0>>1$ pair production occurs in the so-called quasi-static or non-perturbative regime, which manifests itself as a quantum "tunnelling" of an electron from the negative-energy Dirac sea to the positive energy levels. The non-perturbative nature of the process in this regime results in an exponential suppression of the tunnelling rate $R\sim\exp(-8/3\kappa)$, with $\kappa=(2\hbar\omega_\gamma/m_e c^2)(E/E_{crit})$ [4].

It should be noted that one previous experiment has been performed to investigate the Breit-Wheeler reaction experimentally on SLAC[3]. In contrast to our proposal, the SLAC experiment was *not* performed as a pure vacuum interaction, instead the laser interacted with a 46 GeV electron beam. Consequently there were two reactions that could result in pair production in the SLAC experiment. Either a two-step Breit-Wheeler process, whereby the relativistic electron interacts with the laser field to backscatter an γ-ray photon $e+N\omega_{OPT} \Rightarrow e' + \omega_\gamma$ followed by the Breit-Wheeler reaction $\omega_\gamma+N\omega_{OPT} \Rightarrow e^+e$ or via the direct Trident reaction $e + N\omega_{OPT} \Rightarrow e' + e^+e^-$. The SLAC experiment observed pair-production at a low level which was attributed to the two-step reaction. However, more recent theoretical work by Bell and Kirk [6] and others [7,8] suggests that for the parameters region of the SLAC experiment the two processes may be hard to distinguish both theoretically and experimentally and that the number of pairs observed could be actually consistent with the Trident process. Consequently an experiment with a pure γ-optical interaction in a vacuum is required to provide an unambiguous experimental observation of the Breit-Wheeler reaction and hence the first unambiguous observation of pair production from pure light.

The unique combination of two PW class beams at Astra Gemini allow a GeV γ-ray beam to be produced via the efficient conversion of GeV laser accelerated electrons [9] in an optimised tantalum converter with one beam which can then interact in an ultra-high vacuum with the second 0.5 PW Gemini beam (which is expected to reach a nominal peak intensity of $3 \times 10^{22}$ Wcm$^{-2}$ using an f/1 with adaptive optics [5], corresponding to $a_0 \sim 10^2$). Our calculations based on Ref. [4] show that for realistic γ-ray beams with a photon energy of 1GeV the number of pairs created in one laser shot of 40 fs exceeds unity (Fig. 1). These parameters not only allow the first experiments of pair production in a vacuum, but also the first experimental investigation of the non-perturbative, tunnelling regime of vacuum pair-production.

The proposed experimental configuration is also ideally suited to the investigation of the interaction of intense fields with GeV electrons and hence a wide range of strong field QED effects that are hitherto unexplored and these will be the focus of the initial experiments in our programme. For example the proposed experiments offer the opportunity to investigate effects such as the pair-production from electron-laser interactions with rates in excess of 100 pairs per shot possible [8] and to enter for the first time the so-called Quantum Radiation Dominated Regime (QRDR) [10]. This regime is realized at electron energies ε and counter-propagating laser parameters such that the conditions $\chi=(2\varepsilon/m_ec^2)(E/E_{crit})\approx1$ and $R_Q=\alpha a_0\approx1$ are fulfilled, with $\alpha \approx 1/137$ being the fine-structure constant (for an electron with energy ε=1 GeV and a laser field with intensity $3 \times 10^{22}$ Wcm$^{-2}$ it is $\chi \approx 1.4$ and $R_Q \approx 0.9$). In the QRDR the electron emits more than one γ-ray photon already in one laser period. As a consequence, both quantum electrodynamics effects and radiation-reaction effects due to the interaction of the electron with its self electromagnetic field play a dominant role and strongly affect the electron dynamics and photon emission. The theoretical results in [10] show a significant deviation of the expected γ-ray photon spectrum with respect to "conventional" multi-photon Compton scattering, which takes into account the emission of only one γ-ray photon. The versatility of our experimental setup allows us to address fundamental problems also in the low-energy regime where quantum effects are negligible and classical electrodynamics holds. A fundamental and still unsolved problem in classical electrodynamics is he so-called radiation-reaction problem: what is the equation of motion of an electron in an external electromagnetic background field when the interaction of the electron with its own electromagnetic field (radiation reaction) is also taken into account? Dirac [11] proposed in 1938 an equation, now known as Lorentz-Abraham-Dirac (LAD) equation, which however suffers from inconsistencies like the existence of runaway solutions and pre-acceleration effects. The LAD equation has become one of the most controversial equations of physics. More recently Landau and Lifshitz have proposed a different equation, which is originally derived from the LAD one but which does not suffer from runaway solutions and pre-acceleration effects [12]. Neither the LAD equation nor the Landau-Lifshitz one have been tested experimentally. The theoretical results in [13] show that in a regime in which the initial energy ε of the electron is such that $\varepsilon \approx m_ec^2a_0/2$, the dynamics of the electron is particularly sensitive to radiation reaction. In [13] it is shown that in a setup feasible at Astra Gemini, where a counter-propagating electron beam with energy of the order of 30-40 MeV head-on collides with a laser field with intensity of the order of $10^{22}$ Wcm$^{-2}$, the predictions of the Landau-Lifshitz equation could in principle be tested for the first time. Thus the Astra Gemini laser provides a currently unique experimental opportunity for the first observation of the conversion of field energy to matter in a vacuum and for investigating unexplored extreme radiation regimes.

**Experimental Programme:**

Due to the complexity of the experimental programme it is anticipated that the demonstration of pair production from vacuum and of the QRDR will require 3 separate experimental campaigns in which we develop the key components required for a successful run.

*Experiment 1:*
- *Electron beam optimisation and stability at ~1GeV*
- *γ-ray yield characterisation*
- *Determination of background signal and comparison to MonteCarlo prediction*
- *Non-linear Compton scattering immediately after gas jet (no quadrupole imaging)*

*Experiment 2:*
- *Non-linear Compton scattering below and in the QRDR from laser-GeV electrons at secondary interaction point*
- *Pair-production from electrons*
- *Testing of f/1 high vacuum geometry*

*Experiment 3:*
- *Gamma-optical pair creation in high vacuum (Breit-Wheeler process)*

First experiments on Astra Gemini have shown electron beams with energies approaching a GeV and a charge of ~100pC and this experiment will use a modified version of this tested set-up (fig. 2). These experiments used Gemini in an early stage development and used less energy (~ 5-8 J) and had a longer pulse duration (~60fs) than optimal. Thus there is every confidence that we can at least deliver these parameters or better for our QED experiments. Our team includes the experienced electron acceleration groups from Imperial College and Michigan and we thus have high confidence of delivering this aspect of the experiment. The f/1 parabola that will be needed for future experiments has been purchased by QUB as part of the LIBRA particle acceleration project and will be tested in other experiments before it is needed in our second campaign.

We have performed Monte-Carlo simulations using MCNPX, which show a collimated beam of γ-rays emerging from the Bremsstrahlung converter. Each GeV electron is expected to yield in $4 \times 10^{-7}$ photons/$\mu m^2$ in a 0.8 -1 GeV range at the peak of the distribution shown in Fig. 3. Consequently a beam with $2.5 \times 10^8$ electrons (~40 pC) will produce 100 photons/$\mu m^2$ – enough to observe pair production from vacuum in a single shot. Fig. 1 elucidates that even small improvements in terms of γ-ray energy and peak intensity can improve the expected signal substantially.

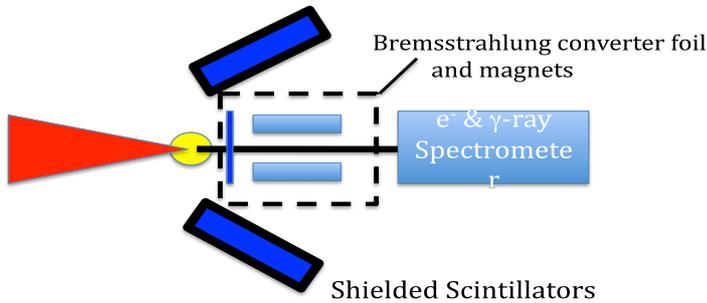

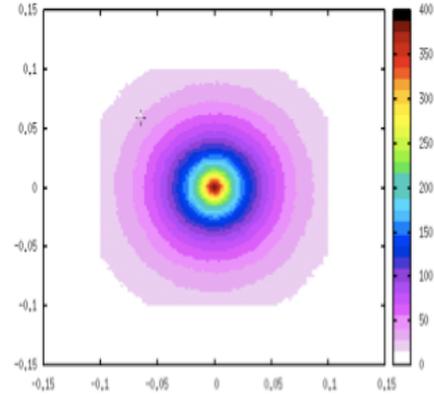

Fig. 2: Sketch of the experimental set-up showing electron spectrometer, γ-ray converter set-up and shielded scintillators to act as test bed for the design of the pair-detectors in later experiments. The data from these test detectors will be compared to MCNPX calculations based on CAD drawings of the Gemini Target Area. In the full experiment the magnets after the converter foil will ensure a pure γ-optical vacuum interaction.

Fig. 3: Intensity distribution of the γ-ray beam in the laser interaction plane ~15cm from the Bremsstrahlung converter. (axes in cm, intensity a.u.)

The electron spectra from the gas target will be detected using the spectrometer/detector set-up proven in previous experiments. The production of GeV electron beams and their stability will be characterised as a function of the laser beam parameters. We will be fielding a recently developed jet target that allows multiple nozzles seperated by ~50µm allowing the target gas to be varied along the direction of laser propagation. Our simulations suggest that this allows highly controlled ionisation injection in a narrow spatial region of the jet and therefore enhanced electron beam control. Once the electron beam has been optimsed and fully characterised, it will impact a high-Z foil (such as Tantalium) in order to generate a γ-ray beam which will be also characterised. The achievable γ-ray flux is in fact of crucial importance to the overall campaign and will be characterised for different foil thicknesses and materials. The emerging γ-ray will be separated from the electrons via a permanent magnet. The γ-ray spectra will be measured using shielded scintillator detectors or by reconverting the γ-rays to pairs in an additional converter foil placed ahead of the electron spectrometer. However, while the pair production rates predicted by theory are in principle easily detected, the challenge will be to minimise the background in the electron positron detectors. We have designed detectors to measure the γ-ray yield using Monte-Carlo radiation codes (as shown in figure 3) and have found designs that are predicted to result in less than 1 background event due to γ-ray/electrons per shot. The signal levels in shielded scintillator detectors which will be placed in the approximate position of the pair detectors in the ultimate experiment will be used to determine real background levels and compare them to our full scale simulations of the Gemini target area using MCNPX. A specially designed beam dump will be fielded to minimise background signal levels.

Note that achieving the required spatial and temporal overlap is within the measured capabilities of the Gemini laser, especially in the case of γ-ray beam which has a significantly larger interaction area than the laser beam. The slight electron beam pointing variations from wake-field acclerators will be controlled by imaging the electron beam to a secondary interaction point using small permanent magnet quadropole lenses. Suitable vacuum conditions will be produced in the final experiment by placing the interaction volume into a secondary chamber with additional pumping within the main Gemini chamber. The scale of the Gemini chamber makes this feasible and the leak rate through the small entrance holes from the Gemini vacuum ($10^{-5}$mbar) to the interaction chamber ($10^{-8}$mbar or 0.1 atoms in the interaction volume) would be negligible[14]. The experimental set-up will use the f/20-f/2 geometry identified in the call for proposals.

**Affiliations:**
1: Max-Planck-Institut für Kernphysik, Heidelberg, 2: Helmholtz Institut, Jena,
3: TEI Crete, 4: Imperial College, London, 5: U. Michigan 6: Queen's Univ. Belfast